\documentclass[
reprint,
superscriptaddress,
bibnotes,
 amsmath,amssymb,
 aps,
prb,
]{revtex4-2}
\usepackage{bm}
\usepackage{xcolor}
\usepackage{graphicx}
\usepackage{lineno}
\usepackage{comment}
\usepackage{dsfont}
\usepackage{physics}
\usepackage{booktabs}
\usepackage{orcidlink}
\usepackage{multirow} 
\usepackage[T1]{fontenc}
\usepackage[caption=false]{subfig} 
\newcommand{\ben}{\begin{equation*}}
\newcommand{\een}{\end{equation*}}
\newcommand{\bean}{\begin{eqnarray*}}
\newcommand{\eean}{\end{eqnarray*}}
\newcommand{\bea}{\begin{eqnarray}}
\newcommand{\eea}{\end{eqnarray}}
\newcommand{\be}{\begin{equation}}
\newcommand{\ee}{\end{equation}}
\newcommand{\rom}[1]{\uppercase\expandafter{\romannumeral #1\relax}}
\def\mos        {MoS$_2$}
\def\Hzero      {\widehat{H}_0}
\def\Hone       {\widehat{H}_1}
\def\Htwo       {\widehat{H}_2}
\def\RR		    {{\bf R}}
\def\rr		    {{\bf r}}

\def\kk         {{\bf k}}
\def\qq         {{\bf q }}
\def\nk         {n\kk}
\def\go         {\omega}
\def\gee        {\varepsilon}
\def\gl         {\lambda}
\def\goql       {\omega_{\qq \gl}}
\def\gql        {\qq \gl}
\def\gsq         {{\mid g^{\gql}_{n n' \kk} \mid}^2}
\def\ga         {\alpha}
\def\gb         {\beta}
\def\la         {\langle}
\def\ra         {\rangle}
\def\Vscf       {\widehat{V}_{scf} }
\def\gS         {\Sigma}
\def\goqlt       {\omega_{\tilde \qq \gl}}
\def\ep         {electron--phonon }
\def\efield	{ {\bf {\cal E}}}
\renewcommand{\[}{\left[}
\renewcommand{\]}{\right]}
\renewcommand{\(}{\left(}
\renewcommand{\)}{\right)}
\begin{document}

\title{Temperature-Dependent nonlinear optics from first-principles: Second-Harmonic Generation in few-layers \mos{}}

\newcommand{\cinam}{CNRS/Aix-Marseille Universit\'e, Centre Interdisciplinaire de Nanoscience de Marseille UMR 7325 Campus de Luminy, 13288 Marseille Cedex 9, France}
\newcommand{\qub}{School of Mathematics and Physics, Queen’s University Belfast, Belfast BT7 1NN, United Kingdom}
\newcommand{\etsf}{European Theoretical Spectroscopy Facilities (ETSF)}
\newcommand{\CLA}[1]{\textcolor{olive}{[PL: #1]}}
\newcommand{\Placeholder}[1]{\textcolor{red}{#1}}

\author{\orcidlink{0009-0009-8692-7686} Anna Romani}
\affiliation{\qub}
\affiliation{\etsf}

\author{\orcidlink{0000-0002-7660-261X} Claudio Attaccalite}
\affiliation{\cinam}
\affiliation{\etsf}

\author{\orcidlink{0000-0002-2549-63} Myrta Gr\"uning}
\email[]{m.gruening@qub.ac.uk}
\affiliation{\qub}
\affiliation{\etsf}

\date{\today}

\begin{abstract}
We present a first-principles real-time approach to study non-linear response of solids at finite temperature. Finite temperature effects are included as renormalization of the quasiparticle energies and a dephasing term proportional to the quasiparticle lifetimes. We evaluate \ep matrix elements from Density-Functional Perturbation Theory for lattice dynamics and then calculate quasiparticles renormalization and lifetime from the Fan and Debye-Waller terms of the electron self-energy. Electron excitations are treated at the independent particle level of approximation. We apply the approach to the second-harmonic generation (SHG) in monolayer and trilayer \mos. We observe a nontrivial temperature-dependence of the SHG due to a strong crystal-momentum dependent quasiparticle renormalization. From the phonon-mode analysis we find that the coupling with acoustic and shear modes determines the overall crystal-momentum dependence respectively in monolayer and trilayer \mos. The nontrivial temperature-dependence of the SHG can help rationalize the increase of SHG intensity with increasing temperature observed in monolayer \mos{} at a given laser energy [Adv. Optical Mater. 8, 2000441 (2020)]. 
\end{abstract}
\maketitle
\section{Introduction}
Thermal effects have a strong impact on nonlinear optical properties. Resonance conditions change with temperature as it contributes both to shifting excitation energies and to broadening. Additionally, phase-matching is also strongly dependent on temperature. Then, thermal effects play a major role in conversion efficiency and operation of photonic devices (see e.g. \cite{PhysRevApplied.11.044084}). Conversely, the sensitivity of nonlinear optical properties to temperature can be used, for example, to characterize the molecular transport in biological systems\cite{hamal2021influence}, to probe valley-population imbalance~\cite{mouchliadis2021probing} and quantum-interference~\cite{lin2019quantum} in transition metal dichalcogenides (TMDs) or to realize compact temperature sensors\cite{8590813}. 

In all these instances, it is key to have an insight into the microscopic mechanisms of the phenomenon and thus it is desirable to have first-principles approaches to run material-specific finite-temperature simulations of nonlinear optics effects. As an illustration, experimental measurements have revealed a remarkable enhancement with increasing temperature of the second harmonic generation (SHG)  of metal dichalcogenides monolayers\cite{khan2020extraordinary}  which calls for a comprehensive treatment of thermal effects and nonlinear optical effects.  

In recent years, several first-principles approaches have been developed to treat electronic and optical properties at finite temperature either based on direct calculation of \ep interaction\cite{giustino2017electron}, on finite difference displacements\cite{monserrat2018electron} or on molecular dynamics\cite{garbuio2009excited}. 
These approaches predict and reproduce band gap renormalization,\cite{antonius2015dynamical,kawai2014electron} finite temperature absorption,\cite{huang2021exciton,alvertis2020impact} phonon-replica in absorption and emission spectra\cite{chen2020exciton}. However, most of finite temperature simulations of optical response have been limited so far to the linear regime, due to the complexity of coupling simulations at finite temperatures with nonlinear response calculations\cite{kawai2014electron}.

Here, we present a first-principles formulation of nonlinear optical properties at finite temperature in dielectrics\footnote{Since we only considering finite-gap materials no temperature effects are considered for the electronic part of the Hamiltonian.}. We combine the approach presented in Refs.~\cite{attaccalite2013nonlinear,Myrta2014}---which allows one to obtain nonlinear susceptibilities of finite-gap periodic systems from real-time simulations---with the Many-Body Perturbation Theory (MBPT) for the \ep coupling\cite{giustino2017electron}---from which we obtain the quasiparticle renormalization and the dephasing factors (Sec.~\ref{sec:methods}). 

We apply the approach to calculate the finite temperature second harmonic generation (SHG) of the mono- and trilayer of \mos{} (Sec.~\ref{sec:results}). These systems are of particular interest considering recent experiments~\cite{mouchliadis2021probing,lin2019quantum,khan2020extraordinary} on TMDs, especially the observation of the SHG intensity enhancement with increasing temperature in TMDs monolayer (in contrast with what observed in 3,5 and 7 layers). From first-principles calculations, the enhancement in the monolayers was attributed to the temperature-induced expansion of the metal-chalcogen bond, whereas the reduction in the few-layers samples was attributed to the increased interlayer distance. These calculations {\em did not consider effects from the \ep interactions}. Besides demonstrating the proposed approach, we address the question of the origin of the SHG increase with temperature in monolayer \mos{}. The objectives are to describe how spectral features are affected by \ep interaction and to analyze what the relevant mechanisms are at microscopic level.

Our results show that for both mono- and trilayer, including the \ep coupling has a strong impact on the main spectral features of the second-harmonic (Sec.~\ref{sec:spectra}). Overall, we see that the \ep coupling affects the SHG spectrum in a nontrivial way and though the increase of temperature lead to an overall decrease in the intensity due to the increased scattering broadening, there are spectral regions where the SHG increases with increasing temperature due to the large QP corrections at van-Hove peak-singularities (Sec.~\ref{sec:qpcorr}). The analysis of the phonon modes (Sec.~\ref{sec:phonon}) shows the prominent role of the coupling with acoustic modes in the monolayer and with shear modes in the trilayer. The strong crystal-momentum dependence of these modes provides a rationale for the nontrivial temperature-dependence of the SHG spectra (Sec.~\ref{sec:discussion}).             

\section{Methods}\label{sec:methods}
T-dependent SHG is obtained combining the calculation of \ep coupling within Many-Body Perturbation Theory (MBPT) with the real-time dynamics of Bloch electrons. We briefly summarise the main ingredients of the approaches in Secs.~\ref{sec:MBPT} and~\ref{sec:realT}, referring to the original implementation for more details. We show then how the equation-of-motions for the real-time dynamics of Bloch electrons are modified to include both the dephasing and the quasiparticle renormalization that stems from the \ep coupling (Sec.~\ref{sec:deph}). We prove that, in the linear-response limit, the proposed approach coincides with the usual approach to calculate finite-temperature optical absorption (Sec.~\ref{sec:linlim}). Finally, we discuss the limitations of the approach (Sec.~\ref{sec:limapp}) and report the computational details (Sec~\ref{sec:compdet}).
\subsection{Electron--phonon coupling}
\label{sec:MBPT}
The total Hamiltonian of the coupled electron--nuclei system $\hat{H}$ can be divided into three parts
\begin{align}
\widehat{H} = \Hzero + \Hone + \Htwo,
\label{eq:sec_theory_1}
\end{align}
where  $\hat{H_0}$ is the electronic Hamiltonian corresponding to the case where the atoms are frozen at their equilibrium positions $\RR_0$---in our case this is the Kohn-Sham Hamiltonian---$\Hone$ and $\Htwo$ represent, respectively, the first and second term in the Taylor expansion of $\Hzero$ when the atomic positions $\{\RR\}$ are expanded around the equilibrium positions $\{\RR_0\}$.

Within MBPT~\cite{kawai2014electron} the exact single particle excitation energies of the total Hamiltonian $\widehat{H}$
are obtained as poles of the Green's Function $G_{\nk}(\go)$ that is solution
of the Dyson Equation:
\begin{equation}
    G_{\nk}\(\go\)=G^{\(0\)}_{\nk}(\go)\[1+\Sigma_{\nk}\(\go\)G_{\nk}\(\go\)\],
\label{eq:sec_theory_2a}
\end{equation}
where the self-energy $\Sigma$ is calculated in terms of $\Hone$ and $\Htwo$.
We consider the two lowest-order nonvanishing contributions to $\Sigma$ (written as functionals of the noninteracting Green's function $G^{0}_{\nk}(\go)$), the so-called Fan and Debye--Waller\,(DW) terms.

The Fan contribution~\cite{fan1950} arises from the second--order term in the perturbative expansion in powers of $\Hone$ gives  to the self-energy
\begin{align}
&\Sigma^{Fan}_{\nk}\(\go,T\) = \sum_{n'\gql} \frac {\gsq}{N_q}
\[ \frac{N_{\qq\gl}\(T\)+1-f_{n'\kk-\qq}}{\go-\gee_{n' \kk-\qq} -\goql -i0^{+}} \right.  \notag \\ &+ 
 \left. \frac{N_{\qq\gl}\(T\)+f_{n' \kk-\qq}}{\go-\gee_{n' \kk-\qq}+\goql -i0^{+}}\],
\label{eq:Fan}
\end{align}
where $\gee_{n' \kk-\qq}$ is Kohn--Sham energy of the $n'$th band at the point $\kk-\qq$ in the Brillouin zone.
$\goql$ is phonon energy relative to the mode $\lambda$ and transferred momentum $\qq$.
$N_{\qq\gl}\(T\)$ is the Bose--Einstein distribution function of the phonon mode $\(\qq,\gl\)$ at temperature $T$
and $f_{n' \kk-\qq}$ is the occupation number of the bare electronic state at $\(n', \kk-\qq\)$.
$g^{\gql}_{n' n \kk}$ are the electron--phonon matrix element~\cite{kawai2014electron} defined as:
\begin{align}\label{eq:gkkp}
g^{\gql}_{n n' \kk}=\sum_{s \ga} \(2 M_s \go_{\gql}\)^{-1/2} e^{i\qq\cdot\tau_s}  \times \\ \la n\kk |\frac{\partial \Vscf\(\rr\)}{\partial{R_{s\ga}}} | n' \kk-\qq \ra \xi_{\ga}\(\qq \gl|s\)
 \notag 
\end{align}
with $M_s$ is the mass of the atom which position in the unit cell is $\tau_s$. $\xi_{\ga}\(\qq \gl|s\)$ are the phonon polarization vectors. All ingredients of Eq.~\eqref{eq:gkkp} are calculated by using DFPT.\cite{baroni2001phonons}

The DW contribution arises from the first-order term in the perturbative expansion in powers of $\Htwo$,
\begin{align}
\gS^{DW}_{\nk}\(T\)=\frac{1}{N_q}\sum_{\gql} \Lambda^{\qq\gl,-\qq\gl}_{n n \kk}  \(2 N_{\qq\gl}\(T\) +1\),
\label{eq:sec_theory_9}
\end{align}
where $\Lambda^{\qq\gl,-\qq\gl}_{n n \kk}$ is a second-order \ep matrix element~\cite{kawai2014electron}:
\begin{align}
    \Lambda^{\gql,\qq'\gl'}_{n n' \kk}= \frac{1}{2}\sum_{s}\sum_{\ga,\gb} \frac{ \xi^{*}_{\ga}\(\qq \gl|s\)
    \xi_{\gb}\(\qq' \gl'|s\)}
{2M_s\(\go_{\gql} \go_{\qq' \gl'} \)^{1/2}} \notag \\
\times \la n\kk |\frac{\partial^2 \Vscf\(\rr\)}{\partial{R_{s\ga}}\partial{R_{s\gb}}} | n' \kk-\qq-\qq' \ra.
\label{eq:lambda_factors}
\end{align}

By solving explicitly Eq.(\ref{eq:sec_theory_2a})
the fully interacting Green's function $G_{nk}\(\go,T\)$ can be written as
\begin{equation}
G_{\nk}\(\go,T\)=\frac {1}{\go-\gee_{\nk}-\gS^{Fan}_{\nk}\(\go,T\)-\gS^{DW}_{\nk}\(T\)}.
\label{eq:Dyson}
\end{equation}
The imaginary part of the Green's function $A_{\nk}\(\go,T\)\equiv\pi^{-1}\mid\Im\[G_{\nk}\(\go,T\)\]\mid$
gives the electronic spectral function. By expanding the $\omega$-dependence of the self-energy around the bare electronic energy, the pole of  $G_{nk}\(\go,T\)$, $E_{\nk}(T)$ is given by
\begin{equation}
E_{\nk}\(T\) = \gee_{\nk} + Z_{\nk}\(T\) \[\gS^{Fan}_{\nk}\(\gee_{\nk},T\)+\Sigma^{DW}_{\nk}\(T\)\],
\label{eq:QP_energy}
\end{equation}
where $Z_{\nk}\(T\) = \(1-\left. \frac{\partial \gS^{Fan}_{\nk}\(\go,T\)}{\partial \go}\right|_{\go=\gee_{\nk}}\)^{-1}$ 
is the renormalization factor.

The Fan and DW self-energies are complex and real functions respectively. Then, the former gives both an \ep induced energy shift and broadening while the latter contributes only with a constant energy shift. Both self-energies depend explicitly on the temperature $T$ via the $N_{\qq\gl}\(T\)$ factor.

\subsection{Nonlinear response from real-time dynamics}\label{sec:realT}
To calculate the nonlinear response we first integrate in time the equation on motion for the Bloch electrons in an external time-dependent electric field. This set of coupled single particle effective time-dependent Schr\"odinger equations reads as,
\be
i\hbar  \frac{d}{dt}| v_{m\kk} \ra = 
\[ \hat H_\kk + \hat W_\kk (\efield) \] 
| v_{m\kk} \ra, \label{eom}
\ee
where $\hat H_\kk$ is the effective single particle Hamiltonian  and $| v_{\kk,m} \rangle$ are the Bloch electrons states. The Hermitian operator $\hat W_\kk(\efield)$ 
represents the coupling with the field $\efield$ consistent with the dynamical Berry-phase polarization ${\bf P}(t)$,\cite{attaccalite2013nonlinear,souza2004dynamics} (see also Appendix~\ref{linearP}), with $\hat W_\kk(\efield) \ket {v_{m\kk}} \propto \efield \cdot \frac{i \partial}{\partial_\kk} \ket {v_{m\kk}} $.
From the solutions of Eq.~\eqref{eom}, the dynamic polarization is calculated as a Berry phase\cite{attaccalite2013nonlinear,souza2004dynamics} from which we extract (non)linear response functions using Fourier analysis\cite{10.21468/SciPostPhys.19.5.129}. The level of theory of the effective Hamiltonian $\hat H_\kk$ corresponds to different approximations for the (non)linear response functions, from the independent particle to the screened Hartree-Fock, see Ref.~\cite{attaccalite2013nonlinear}. The Bloch-electron states are written in the basis of Kohn--Sham states $\{| u_{\kk,i} \rangle\}$ of the unperturbed system,
\be
| v_{n\kk}(t)\rangle = \sum_{i=1}^{\infty} c_{ni\kk}(t) | u_{i\kk} \rangle,
\ee
so that 
then Eq.~\ref{eom} becomes an equation for the coefficients $c_{ni\kk}(t)$,
\be
i\hbar  \frac{d}{dt}  c_{nj\kk}(t)  = \left[\hat H_\kk + \hat W_\kk(\efield)\right]_{j,i}  c_{ni\kk}(t). \label{eom_ks_basis}
\ee

\subsection{Phonon-induced dephasing and quasiparticle renormalization}
\label{sec:deph}
The Hamiltonian in Eq.~\eqref{eom} is Hermitian and does not include any dissipative term. In practical calculation a phenomenological dephasing is added either during post-processing\cite{octopus} or directly in the Bloch-electron dynamics\cite{attaccalite2013nonlinear}. Here, we introduce in Eq.~\eqref{eom} a dephasing term that depends on the temperature which is derived from the imaginary part of the Fan self-energy in Eq.~\eqref{eq:Fan}. 
From the Bloch-electron states we can define the density matrix as,
\be
\hat \rho_{\kk} (t) =  f \sum_{m=1}^{N_v}  | v_{m\kk} (t) \rangle \langle v_{m\kk} (t) |
\ee
where $f$ is the spin degeneracy and the sum runs on all $N_v$ occupied bands.
In the Kohn--Sham basis, 
\be
\rho_{ij\kk} (t) =  f \sum_{m=1}^{N_v}  c_{mj\kk}^* (t) c_{mi\kk} (t).
\label{eq:densmat}
\ee
Using the time-dependent density matrix [Eq.~\eqref{eq:densmat}] we then define the dephasing operator as a generalization of the phenomenological dephasing used in Ref.~\cite{attaccalite2013nonlinear,skachkov2026linear},
\bea
\Gamma_{ij\kk} &\equiv&-i \gamma_{ij\kk}\left( \rho_{ij\kk} -\rho^0_{ij\kk}  \right) \label{deph_new},
\eea
where $\rho^0_{ij\kk}$ is the unperturbed (zero-field) density-matrix and the $\gamma_{ij\kk}$ are calculated as,
\bea
\gamma_{ij\kk} =\Im{\Sigma^{Fan}_{i \kk}\(\epsilon_{i\kk},T\)}+\Im{\Sigma^{Fan}_{j \kk}\(\epsilon_{j\kk},T\)}. \label{deph_imgsigma}
\eea
This definition includes both the intraband relaxation and carriers recombination~\cite{haug2009quantum}.

Introducing $T$-dependence in the dynamics of the Bloch electrons, amounts simply to replace the effective Hamiltonian in Eq.~\eqref{eom} with the non-Hermitian operator
\be\label{eq:Hop}
\hat { H}_\kk \leftarrow \hat {H}_\kk + \hat{\Gamma}_\kk (T) + \hat{\Delta}_\kk (T)
\ee
where the $\hat{\Delta}_\kk$ operator introduces the quasiparticle energy renormalization in Eq.~\eqref{eq:QP_energy},
\begin{equation}
\hat{\Delta}_\kk = \sum_j (E_{j\kk} -\gee_{j\kk}) | u_{j\kk} \rangle \langle u_{j\kk}|.     
\end{equation}
The approach has been implemented into the {\tt lumen} code~\cite{Sangalli2026-fy}. 

\subsection{The linear-response limit}\label{sec:linlim}
Within the linear response limit, the time-dependent Berry-phase polarization along direction $\alpha$ reads,
\begin{equation}\label{eq:P_lin} 
P^{(1)}_\alpha(t)=-\frac{1}{(2\pi)^3}\sum_{v,c}\int_{BZ}d\kk~2Re[\rho_{cv\kk}^{(1)}(t) d^\alpha_{cv\kk}],
\end{equation}
as derived in Appendix~\ref{linearP} (in what follows we assume an insulating system where $v,v'$ indicate valence bands---fully occupied---and $c,c'$ conduction bands---empty). 
Starting from Eq.~\eqref{eq:P_lin}, we will show that, in the linear-response limit, at the independent (quasi)particle level, the temperature-dependent response function (in frequency-space) corresponding to the dynamics under the non-Hermitian operator in Eq.~\eqref{eq:Hop} has the form,
\begin{align}\label{eq:chi1_fin} 
	&\chi^{(1)}_\alpha(\omega, T)=\notag \\&\int_\text{BZ}~d\kk \sum_{v,c} \frac{|d^\alpha_{vc\kk}|^2}{\omega -(E_{c\kk}(T) - E_{v\kk}(T))+i\gamma_{vc\kk}(T) } + (c 	\leftrightarrow v),
\end{align}
known from the literature~\cite{giustino2017electron}.

In Eq.~\eqref{eq:P_lin}, $\rho^{(1)}_{vc\kk}$ is the variation of the density matrix element with respect to the field-coupling operator,
\begin{equation}\label{eq:rholin}
    \rho^{(1)}_{vc\kk}(t) = \int dt' \sum_{v',c'} R_{vc,v'c'\kk} (t-t') U^{(1)}_{v'c'\kk} (t'),
\end{equation}
where we defined the response function $R$,
\be \label{eq:defchi}
R_{vc,v'c'\kk} (t,t') \equiv \left.\frac{\delta \rho_{vc\kk}(t)}{\delta W_{v'c'\kk}(t')}\right|_{\efield = 0},
\ee
and $U_{v'c\kk}$ is the representation in Kohn-Sham basis of $\hat W_\kk$.
To get an expression for $R$ in terms of microscopic quantities, from the equation of motion for the Bloch-electron dynamics~\footnote{for simplicity of notation we absorb the diagonal quasiparticle correction operator $\hat \Delta_\kk$ into the definition of $\hat H_\kk$: the resulting operator is diagonal in the Kohn-Sham basis with eigenvalues $E_{n\kk}$.},
\be
i|\dot v_{\kk m} \rangle = \left(\hat H_\kk + \hat \Gamma_\kk+ \hat W_\kk \right)|v_{\kk m} \rangle, \label{eom_lr}
\ee
we derive the corresponding equation of motion for the dynamics of the valence-conduction density matrix elements,
\begin{align}\label{eq:eomrho}
i\dot\rho_{vc\kk}= \left [\hat H_\kk + \hat W_\kk, \hat \rho_\kk \right]_{vc} + \left \{\hat \Gamma_\kk, \hat \rho_\kk \right \}_{vc},
\end{align}
where $[\,]$ and $\{\,\}$ are respectively the commutator and anti-commutator operators. 
By taking the functional derivative of Eq.~\eqref{eq:eomrho} with respect $U_{v'c'\kk}$,
we get,
\begin{align} \label{eq:R_t}
	i\partial_t R_{vc,v'c'\kk}(t,t')&=  
    R_{vc,v'c'\kk}(t,t') \left[(E_{c\kk} - E_{v\kk}) \right. \notag \\ \left.+\delta_{vv'}\delta_{cc'} \delta(t-t')    - i\gamma_{vc,\kk}  \}\right].
\end{align} 
Taking the Fourier transform of Eq.~\eqref{eq:R_t}, we obtain an expression for $R$ as a function of the Bloch electron energies $\{\epsilon_i\}$ and \ep coupling $\gamma_{ij\kk}$,
\begin{equation}\label{eq:expr_R} 
	R_{vc,v'c',\kk }(\omega, T)=\dfrac{\delta_{vv', cc´}}{\omega -(E_{\kk c}(T) - E_{\kk v}(T))+i\gamma_{\kk vc}(T) }, 
\end{equation}
where we made explicit the dependence of $T$. Using Eq.~\eqref{eq:expr_R} into the Fourier transform of Eq.~\eqref{eq:rholin}, we obtain,
\begin{equation}\label{eq:rholin2}
    \rho^{(1)}_{vc\kk}(\omega,T) = \efield\dfrac{d_{vc\kk}}{\omega -(E_{\kk c} - E_{\kk v})+i\gamma_{\kk vc}(T) },
\end{equation}
where we used that $U^{(1)}_{vc,\kk} = \efield d_{vc\kk}$ (see Appendix~\ref{linearP}).
Finally, inserting the latter expression into the Fourier transform of Eq.~\eqref{eq:P_lin} and using the constitutive relation between polarization and electric field (Eq.~\eqref{eq:chiE} in Appendix~\ref{linearP}) results in the expression for the finite-temperature linear-response in Eq.~\eqref{eq:chi1_fin}.
This derivation, showing that in the linear-response limit we recover the finite-temperature linear-response expression, can be extended to higher-order response functions, but it is restricted to the independent (quasi)particle approximation (see next Section).

\subsection{Limitations of the approach}\label{sec:limapp}
The proposed approach has three main limitations: it assumes harmonic behavior of the atoms, it does not include phonon-assisted process and it is restricted to the independent-particle treatment of the Bloch-electrons.
The first limitation stems from the underlying DFPT approach which uses the harmonic approximation. The second limitation is due to the static approximation in Eq.~\eqref{deph_imgsigma} as the dynamic self-energy would be needed to describe phonon-assisted processes. The third limitation is less evident and it is ultimately due to treating the \ep lifetime term as a dephasing operator rather than a self-energy~\cite{stefanucci2013nonequilibrium}. In fact, many-body effects (typically excitonic and local field effects) would enter as a {\em 
static} self-energy operator which is a functional of the density matrix. It has been shown~\cite{sangalli2021excitons,antonius2017theory} that the inconsistent treatment of the lifetime operator in the equation of motion leads to an unphysical lifetime of the lowest excitonic states. Such unphysical lifetime is expected to be small in system with small excitonic binding energy but could be severe otherwise. {\em All results in this work have been obtained within the independent-(quasi)particle approximation}. 

\subsection{Computational details}\label{sec:compdet}
The electronic structure, the structural optimization and the \ep matrix elements calculations were performed using the plane-wave DFT code {\tt Quantum Espresso}~\cite{pwscf} at the local density approximation level~\footnote{While generalized gradient approximations give a better estimate of the lattice parameters, we found that for the trilayer, the structure was not dynamical stable preventing phonon calculations likely due to missing van der Waals interactions. Since LDA did not present the issue, for the sake of simplicity we opted for this level of theory rather than attempt the addition of van der Waals corrections.}. We used the optimized norm-conserving Vanderbilt scalar-relativistic pseudopotentials~\cite{PhysRevB.88.085117} from the {\tt pseudodojo} database (v.3.3)~\cite{PseudoDojo} with a kinetic energy cutoff for wavefunctions of 90~Ry. Our calculations do not include spin-orbit coupling effects.  
Phonon modes and frequencies were calculated using the double grid integration in Appendix\ref{sec:double grid} with a $18 \times 18 \times 1$ coarse $\kk$-grid and a  $60  \times 60 \times 1$ fine $\kk$-grid; 150 intermediate states are included in the summation of Eq.~\eqref{eq:Fan}.
The quasiparticle (QP) properties and optical spectra calculations were performed using the {\tt lumen} code~\cite{yambo, Sangalli2026-fy}. The QP energies and lifetimes were calculated for temperatures ranging from 0~K to 500~K and interpolated onto the $\kk$-point grid used in the real-time simulations. The optical susceptibilities were obtained via Fourier analysis of the real-time polarization, as described in Ref.~\cite{10.21468/SciPostPhys.19.5.129}.
Results analysis and graphical representation were performed using {\tt yambopy}~\cite{YamboPy}. Relevant numerical parameters are listed in Table~\ref{tab:convpar}, including the dimension of the supercell in the non-periodic direction, $L_z$, and the effective thickness, $d_\text{eff}$ of the system in that direction. These parameters are used in the definition of the surface second harmonic generation, $\chi_\text{2D}^{(2)} \equiv \chi^{(2)} \times L_z$, and of the {\em effective} imaginary part of the dielectric tensor, $\epsilon_2^\text{eff} \equiv \epsilon_2 \times {L_z/d_\text{eff}}$ where $ \chi^{(2)}$ and $\epsilon_2$ are the bulk quantities from the supercell calculations. The effective thickness, $d_\text{eff}$, has been calculated from the distance between the outer-layer-chalcogen planes plus twice the van der Waals radius~\cite{Klimova2026Materials,surfur_radius}. 
\begin{table}[ht]
\footnotesize
\centering
	\caption{All the parameters used in the nonlinear response calculations for both the monolayer (ML) and trilayer (TL) of \mos{}: the $\kk$-point sampling used in the linear-response and real-time calculations (phonon and quasiparticle in parenthesis), the range of bands considered  and the index of the highest valence band (HVB) in parentheses, the lattice parameter ($a_\text{latt}$), supercell dimension in the non-periodic direction, $L_z$ and the effective thickness, $d_\text{eff}$. 
    }
\begin{tabular}{|c|c|c|c|c|c|}
	       \hline
        System & $\mathbf{k}$-points &  $N_\mathrm{b}$ (HVB) & $a_\text{latt}$ (nm) &$L_z$ (nm) & $d_\text{eff}$ (nm) \\
        \hline
        ML  & $30\times30$  & 9-19 (13)& 0.314 &2.12 &  0.687 \\
        \hline
	      TL  & $30\times30$  & 35-45 (39)& 0.315 & 3.18 &  1.889 \\ 
          \hline
	\end{tabular}
    \label{tab:convpar}
\end{table}

\begin{figure}[ht]
    \centering
    \includegraphics[width=0.95\linewidth]{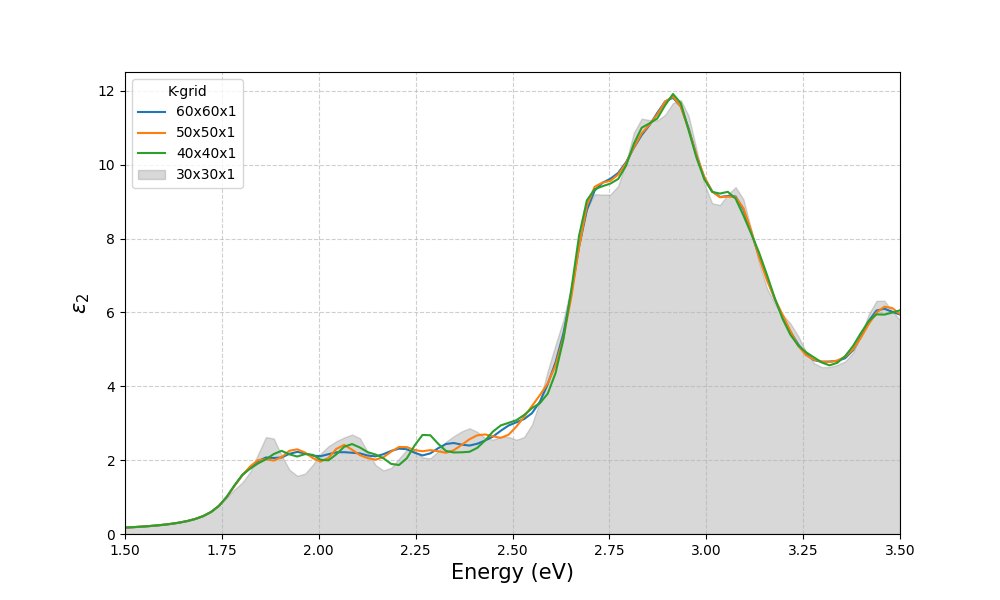}
    \caption{Convergence of the absorption spectrum with the $\kk$-point grid for the monolayer of \mos{} (without temperature effects). The $30\times 30$ (gray area) used in the optical spectra calculations shows spurious peaks below 2.5~eV. Above 2.5~eV, main spectral features are converged with the $30\times 30$ $\kk$-point grid.}
    \label{fig:kpoint_conv}
\end{figure}

As shown in Fig.~\ref{fig:kpoint_conv}, a significantly denser $\kk$-point grid would be required for full convergence of the plateau between 1.75--2.50~eV. In particular, with the $30 \times 30$ $\kk$-point grid used in the calculations, spurious peaks are still visible in this region. In real-time simulations, the applied electric field (in our case the electric field is directed in plane along $xy$) greatly reduces the unit cell symmetries so that a $30 \times 30$ $\kk$-point grid already results in large number of states (900 k-points $\times N_b$), which limits the possibility of using larger grids. In the analysis of the results below, we acknowledge this limitation discussing energy-ranges rather than peaks. Above 2.5~eV, spectral features are well-converged with the $30\times 30$ $\kk$-point grid, including the strong peak at about 2.9~eV and the shoulders at about 2.7~eV and 3.1~eV.

\section{Results} \label{sec:results}
The approach from Sec.~\ref{sec:methods} is applied to the temperature-dependence of the SHG for the mono- and trilayer \mos{} (Sec.~\ref{sec:spectra}). The temperature behavior of the spectra is explained from the analysis of the quasiparticle corrections (Sec.~\ref{sec:qpcorr}) and its decomposition into phonon modes contribution for the relevant electron states and phonon modes (Sec.~\ref{sec:phonon}). The results are then compared with existing experimental results (Sec.~\ref{sec:discussion}).

\subsection{Finite-temperature SHG spectra}\label{sec:spectra}
\begin{figure*}[ht]
\centering
\includegraphics[width=0.9\linewidth]{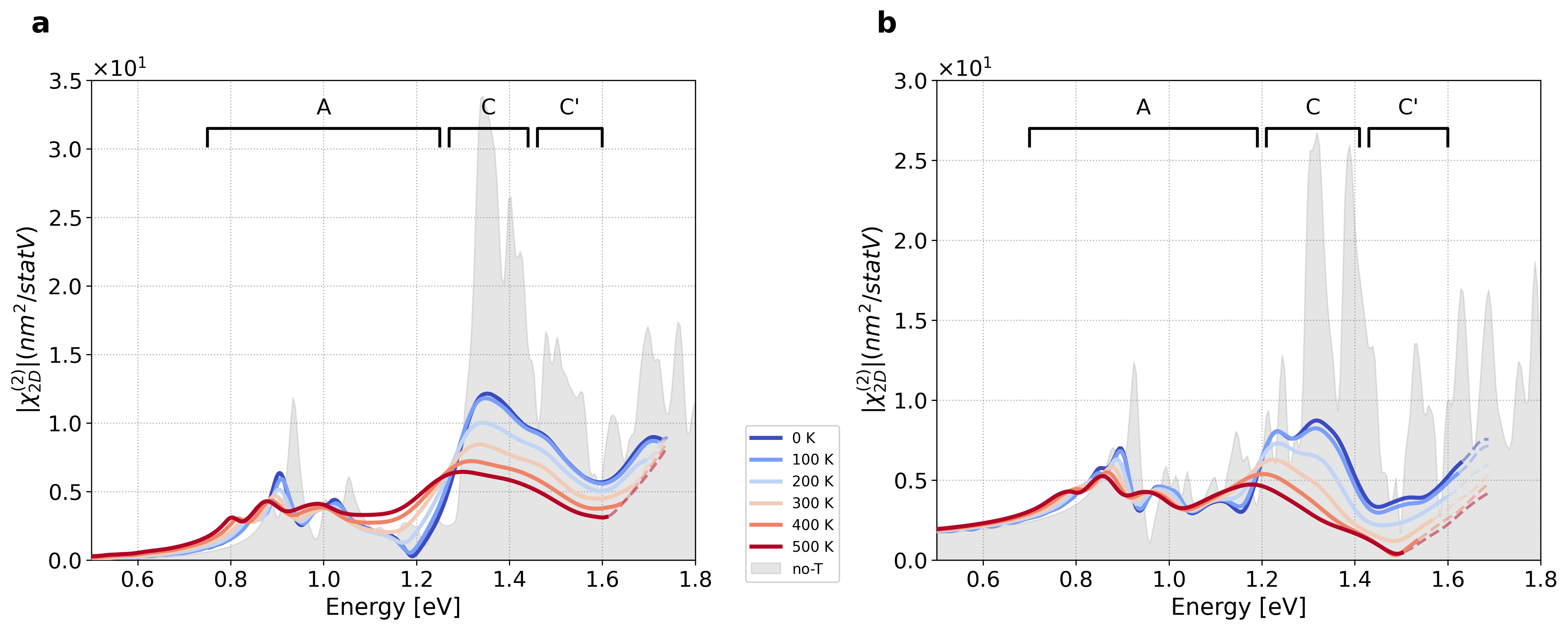} 
    \caption{Absolute value of surface SHG of monolayer (a) and trilayer (b) \mos{} as a function of the laser frequency. Spectra for temperature in the 0-500~K range (blue-to-red lines) are compared with the spectrum at the no-T level  (gray shade) with a phenomenological broadening of 0.02 eV. The otherwise continuous lines of the T-dependent spectra become dashed after the absorption onset to indicate where one-photon resonances started contributing. 
    }  
    \label{fig:chi_2}
\end{figure*}

The absolute value of the temperature-dependent surface SHG, $|\chi^{(2)}_\text{2D}|$,  for the monolayer and trilayer of \mos{} is shown in Fig.~\ref{fig:chi_2}. For comparison, we also report spectra calculated without \ep effects that we refer to as ``no-T". In what follows, the position of the energy of spectral features refer to the no-T spectra when not otherwise indicated. 
In Fig.~\ref{fig:chi_2} (a), for analysis purposes, three main energy ranges below the absorption onset are considered: 0.8--1.3 eV, 1.3--1.45 eV and 1.45--1.6 eV.
In the 0.8--1.3 eV range, we observe a set of close superimposing peaks with similar intensity [that would form a plateau if using a denser $\kk$-point grid, see Fig.~\ref{fig:kpoint_conv}], a characteristic feature of a step-like Van Hove singularity~\cite{PhysRevB.100.075301, Parravicini2013}. This ``plateau" corresponds to the A/B excitons identified in experiments~\cite{doi:10.1021/nn303973r} and in first-principles calculations at the Bethe-Salpeter equation (BSE) level \cite{Myrta2014,PhysRevB.93.155435}, and will be labeled with A in what follows~\footnote{when spin orbit is included, two peaks are distinguishable in the same energy range}.
In the 1.3--1.45 eV range, the SHG spectrum shows an intense split peak at 1.35 eV 
followed by another distinct features in the 1.45--1.6 eV range. These features corresponds to exciton C and C' identified both in experiments~\cite{doi:10.1021/nn303973r} and in first-principles calculations~\cite{Myrta2014,PhysRevB.93.155435} and will be labeled with C and C' consistently.

As a general trend, with increasing temperature, all peaks undergo broadening and a red-shift. Comparing the $T=0$~K with the no-T level of theory, a rather strong zero-point renormalization~\cite{giustino2017electron} is visible. Looking at the energy ranges identified above, we observe that features in ranges C and C' undergo to a substantial broadening and a pronounced red-shift as temperature rises, while in A intensity-renormalization and redshift are rather weak. We notice that while the main spectral features are clearly reduced in intensity with temperature, the combination of energy shift and broadening can locally increase the SHG as for example is observed around 1.2~eV.  
The SHG spectrum of the trilayer [Fig.~\ref{fig:chi_2}(b)] shows analogous features as the monolayer, a plateau A followed by the split peak C and the feature C'. As in the monolayer, features in C and C' are significantly red-shifted and broadened as temperature increases, whereas for the plateau A temperature effects are weaker. An analogous trend is observed for the temperature-dependent absorption spectra both at the independent-particle (see Supplemental material) and at the Bethe-Salpeter level~\cite{PhysRevB.93.155435}.

\begin{figure*}[ht]
\includegraphics[width=0.9\linewidth]{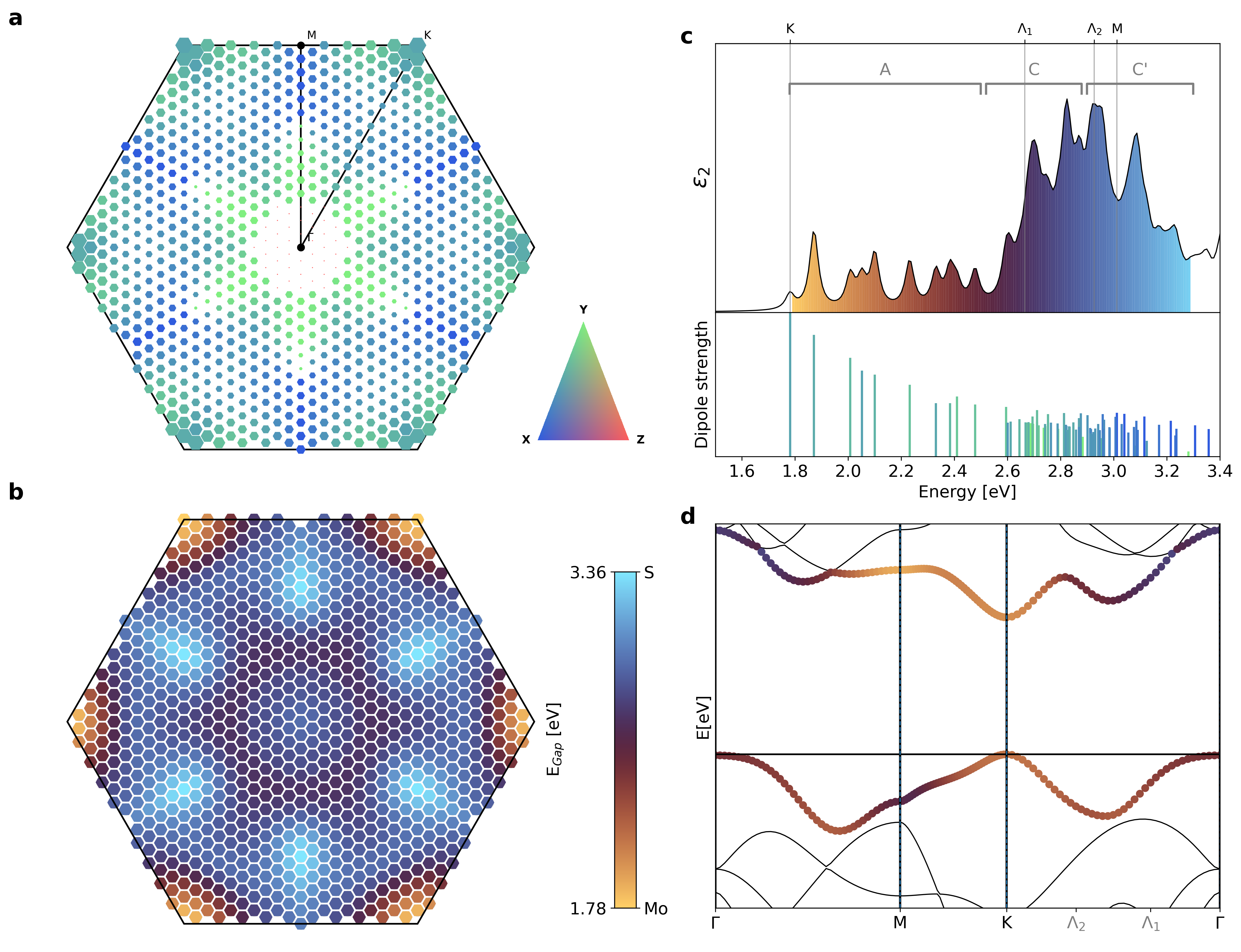}
    \caption{Monolayer \mos{}. (a) Dipole matrix elements $\bf{d}_{cv\kk}$ between the HVB and LCB across the BZ. Marker size gives the strength of the dipole, while the color of the marker gives the contribution of the $x,y,z$ components (blue/green/red triangle). (b) The energy difference $\Delta\epsilon_{vc\kk}$ between LCB and HVB across the BZ.  (c) The absorption spectrum at no-T level. The color indicates the energy difference of the corresponding transition using the same color bar of panel (b) for comparison. The vertical lines indicate the position of the transitions at $K$ and $M$ high-symmetry points and of the $\Lambda_{1,2}$ points (see SM). In the lower panel the dipole strength of the matrix elements in panel (a) are plotted against the corresponding energy gap. (d) Band structure along the $\Gamma-M-K-\Gamma$ path at the LDA level. Fatbands are plotted for the HVB and LCB. The color corresponds to the atomic contribution: yellow for Mo and cyan for S.}
    \label{fig:Mono dipoles analysys}
\end{figure*}

To characterize the spectral features in terms of single-particles transitions, in Figure~\ref{fig:Mono dipoles analysys}, for the monolayer, we analyze the the dipole matrix elements $d_{vc\kk}$ and energies differences $\Delta\epsilon_{vc\kk} = \epsilon_{c\kk} -\epsilon_{v\kk}$ in the Brillouin zone (BZ) [panels (a) and (b)] where $v$ and $c$ are the indexes of the lowest conduction band (LCB) and the highest valence band (HVB). The comparison with the absorption spectrum [panel (c)] at the no-T level allows us to identify the contributions of single-particles transitions.
In panel (a), the strongest dipole elements occur at the $K$ points with in-plane direction. At $M$ and close to $M$ along the $\Sigma$ direction ($M-\Gamma$, see SM) dipoles are oriented along $x$, while farther from $M$, towards $\Gamma$, dipoles are  oriented along $y$. 
Around $\Gamma$, dipoles are oriented out-of-plane and all very weak. In panel (b), the absolute minima for $\Delta\epsilon_{vc\kk}$ occur at the $K$ valleys (1.78 eV) and absolute maxima close to the $M$ points along the $\Sigma$ direction (3.36 eV). Local maxima (hills) are located at $\Gamma$ (~2.8 eV) and along $\Lambda$ (that is $\Gamma-K$) direction close to $K$ (~2.9 eV), while local valleys are located around the $\Gamma$ 'hill' extending towards the $\Lambda$ direction (~2.6 eV). 

In the absorption spectrum [panel (c)], the absolute (local) minima/maxima from panel (b) corresponds to step-like (peak-like) critical points in the spectra~\cite{Parravicini2013}. As discussed in Sec.~\ref{sec:compdet}, the step-like singularity corresponding to the minimum critical point at 1.78 eV is not well visible due to poor convergence with the $\kk$ points, while at 3.36 eV a step-like singularity is well visible. To summarise, the plateau A corresponds to transitions localized at and close to $K$ associated with $xy$-oriented dipoles with significant strength. The features in region C, correspond to transitions from the local valley around the $\Gamma$ 'hill' and extending toward $K$ with a large number of dipoles of comparable strengths oriented in the $xy$ plane and along $y$, while the features in region C' corresponds to the local maxima regions closer to the $K-M$ edge with the dipoles mostly oriented along $x$. Panel (d) shows the band structure close the Fermi energy projected onto the molybdenum and sulfur atoms along the $\Gamma-M-K-\Gamma$ path (a detailed analysis at the high-symmetry point is given in the SM). All transitions have hybrid Mo-S character with a varying degree of hybridization across the BZ. At, and close to, the $K$ points---corresponding to the A region---we have predominantly $d$(Mo)--$d$(Mo) transitions with strong, perfectly isotropic dipoles in the plane.
At $M$ and at the region close to the critical points along $\Sigma$ and $\Gamma$---corresponding to the C/C' regions---transitions from HBV to LCB give a net shift of the electronic density center of mass along the Mo-S bond [$d$(Mo)--$p$(S) or $p$(S)--$d$(Mo)] and the corresponding dipoles acquire a preferred direction either along $x$ or $y$. The same analysis with analogous characterization has been carried out for the trilayer (see SM).

\subsection{Quasiparticle corrections}\label{sec:qpcorr}

\begin{figure*}[ht]
    \centering
    \includegraphics[width=0.9\linewidth]{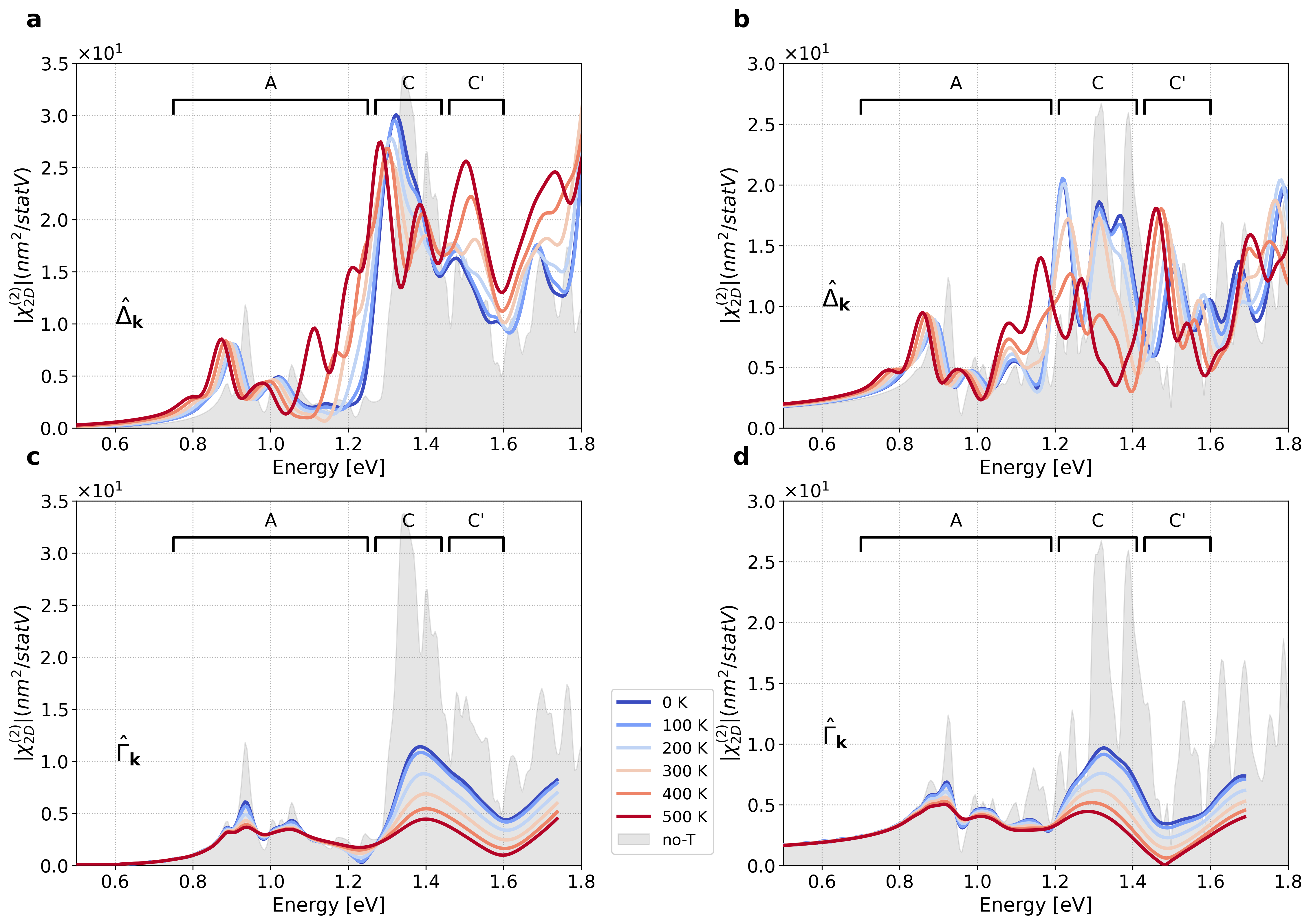}
    \caption{Absolute value of surface SHG of monolayer (a,c) and trilayer (b,d) of \mos{} as a function of the laser frequency obtained at different temperatures (red-to-blue continuous lines) and compared with the no-T level of theory (gray shade). In panel (a) and (b) only the energy renormalization ($\Delta_\kk$) is included in the effective Hamiltonian [Eq.~\eqref{eq:Hop}] while in panel (c) and (d) only the lifetime ($\Gamma_\kk$).}
    \label{fig:chi2 only W only E}
\end{figure*}

First, to disentangle the effects of broadening and energy-shift, we calculate the temperature-dependent SHG with either the quasiparticle shift (only $\Delta_\kk$ in equation ~\ref{eq:Hop}) or the lifetime broadening (only $\Gamma_\kk$ in equation ~\ref{eq:Hop}). Results are shown in Fig.~\ref{fig:chi2 only W only E}.
The difference between regions A and C/C' emerges even more clearly. Both the energy-shift, in panels (a) and (b), and the broadening, in (c) and (d), affect more strongly the higher-energy part of the spectra for both mono- and trilayer. 
Overall, from panels (a) and (b), it is clear that the spectra are {\em not shifted rigidly}. In particular, for the monolayer [panel (a)] at high temperatures some intense transitions contributing to the C peak move to 1.2~eV, which explains the increase with temperature of surface SHG in this region. As well, intense transitions are moving from above the absorption onset to about 1.5~eV~\footnote{Note that while below the absorption onset two-photon transitions are shifted by $\Delta_{mn\kk}/2$, above the onset one-photon transitions are shifted by $\Delta_{mn\kk}$}. Similar large energy shifts of particular transitions are visible in the high-temperature trilayer spectra [panel (b)] . 

\begin{figure*}[ht]
    \centering
\includegraphics[width=0.9\linewidth]{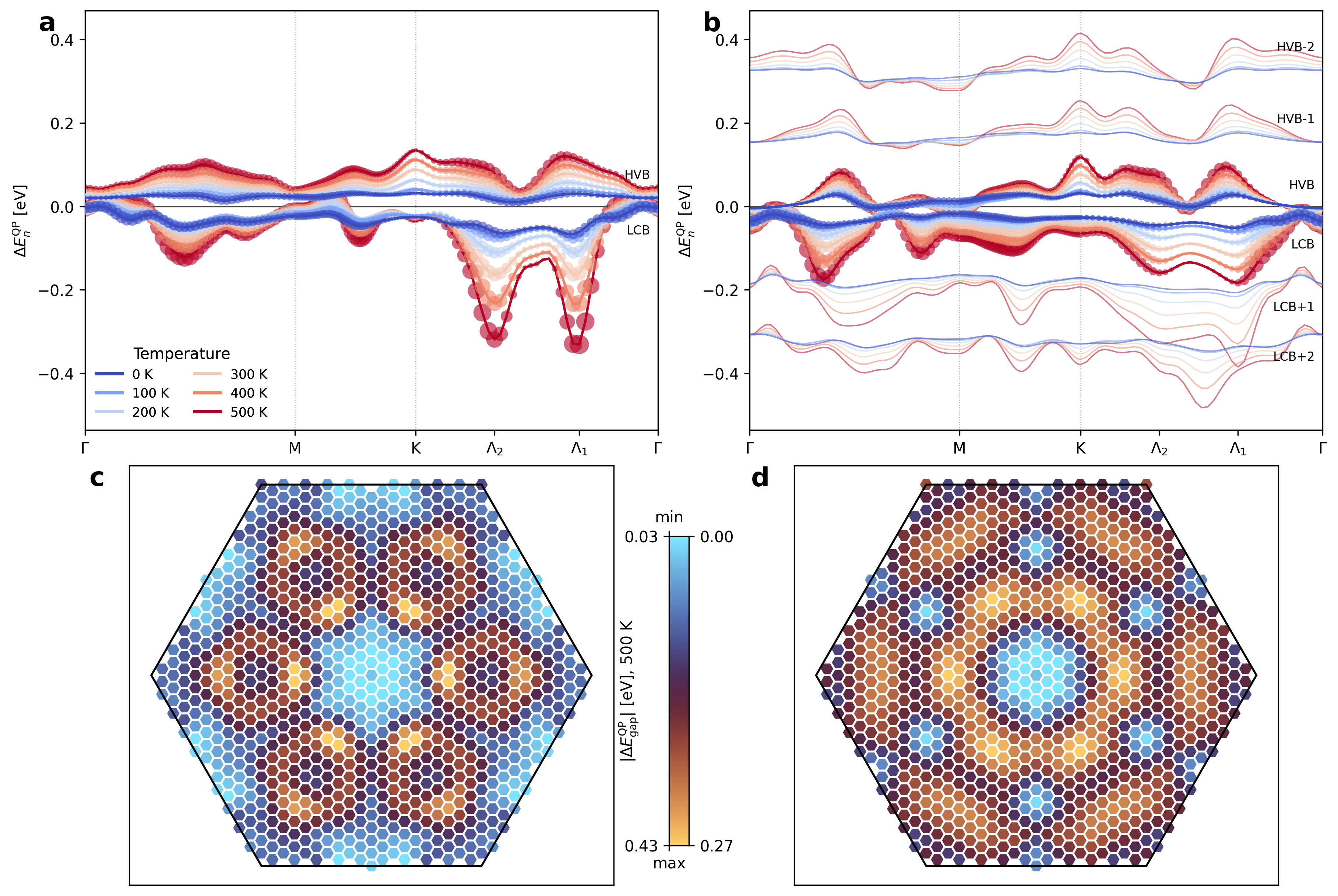}
    \caption{(a),(b): Real part of the quasiparticle correction for monolayer and trilayer \mos{} along the $\Gamma-M-K-\Gamma$ high-symmetry path in the BZ. Energy renormalization of the LCB and the HVB---labeled on the right-$y$-axis---are plotted for temperatures 0, 100, 200, 300, 400, 500 K. The marker size is proportional to the imaginary part of the QP correction, corresponding to the spectral broadening, in eV. For the trilayer, in (b), the the real part of the QP corrections of the near-degenerate bands is shifted by an offset of $\pm$~0.2~eV. For both materials, a strong quasiparticle renormalization is observed at $\Gamma_{1,2}$.
    (c), (d) the quasiparticle renormalization of the bandgap at 500 K across the BZ for monolayer and trilayer \mos{}, respectively. The scale of the QP correction for monolayer and trilayer are shown at the left and right of the color bar.}
        \label{fig:QP_vs_T}
\end{figure*}

Next, in Fig.~\ref{fig:QP_vs_T}, we show the $\kk$-dependence of the relevant quasiparticle corrections (Sec.~\ref{sec:phonon}) that elucidate both the difference between region A and C/C' and the increase of the SHG with temperature observed in some spectral regions.
In panels (a) and (b), for the mono- and trilayer, the QP corrections are analyzed separately for the HVB and LCB along the $\Gamma-M-K-\Gamma$ $\kk$-path.
For both systems, with increasing temperatures the energy shift of the valence bands is mostly positive and that of the conduction bands mostly negative so to result in an overall reduction of the transition energies and consequently red-shift observed in Figs.~\ref{fig:chi_2}. 

For the monolayer [panel (a)], the conduction band is the most affected by the QP corrections and shows a stronger $\kk$-point dependence while the QP corrections for the valence band are overall smaller in absolute value and more uniform along the $\kk$-path. The {\em largest shifts are found in correspondence of the local minima/maxima} as identified in Sec.~\ref{sec:spectra}. In particular, for the conduction band, along $K-\Gamma$, points corresponding to local maxima $\Lambda_{1,2}$, are shifted by up to 0.35 eV at 500K. At the same points, we also observe the largest broadening. 
This is reflected on the QP bandgap renormalization (QP corrections to the HVB $\rightarrow$ LCB transitions) across the BZ, which is shown in panel (c). Corrections are negligible at and around $\Gamma$ and $M$, while show a high variability along $\Lambda$ with the largest corrections (in absolute value since they are negative) corresponding to the $\Lambda_{1,2}$ points previously highlighted. From these results, comparing with Fig.~\ref{fig:Mono dipoles analysys} (b), it is clear that the transitions that contribute to C, even if close in energy at the no-T level, are shifted by very different amounts, explaining what observed for the transitions in the C/C' range [Fig.~\ref{fig:chi2 only W only E} (a)]. In particular, the energy-shift of the two-photon transitions located at the points of largest shift along $\Lambda$ is about 0.2~eV at 500K, and bring these transitions at 1.1-1.2~eV.  Around the $K$ point, the renormalization is much smaller, in agreement with the weak intensity renormalization observed for the A plateau. 

The remarkable QP corrections at the $\Lambda_{1,2}$ points (and their symmetry equivalent in the BZ) can be rationalized by looking at the valley landscape of the BZ and of the LCB in terms of {\em the availability of intermediate electronic states close in energy and the corresponding electronic scattering phase space}~\cite{Lee2021-zo}. In fact, the contribution of each scattering channel to the Fan self-energy [Eq.~\eqref{eq:Fan}] is weighted by the energy denominator involving the initial and intermediate electronic states. The local maxima of the LBC are located close to two minima (at $K$ and at about 1/2 of $\Lambda$ direction), which provide a particularly large scattering phase space. The conduction band around the latter minimum is flatter than around $K$, with consequently, a larger number of intermediate electronic states available.

For the trilayer, Fig.~\ref{fig:QP_vs_T}, the QP correction on LCB and HVB and near-degenerate bands [panel (b)] are affected by energy shift up to 0.2 eV for the highest temperature. Contrary to the monolayer, the broadening along the $\Lambda$ direction for the conduction band is modest and the energy shift reduced. 
The gap renormalization at $500$~K (between the LCB and the HVB) in panel (d) is overall smaller compared to the monolayer. Though the plot is apparently very different, the overall pattern of corrections across the BZ is very similar to the monolayer. Again, largest QP shifts are associated to transitions in the C, C' regions.

To summarize, quasiparticle corrections are the largest at critical saddle points which corresponds to the van-Hove peak-like singularities in the C region. Electrons in states corresponding to these saddle points, stemming from local maxima in the LCB, have a large number of states close in energy to decay to than electrons at and close to the absolute minimum $K$ which corresponds to step-like singularity in the A region. This agrees with the observation in Refs.\cite{PhysRevB.93.155435,tanimura2016formation} that high-energy states renormalize more than states close to the fundamental gap.~\footnote{An electron in an higher conduction state tends to decay in a lower conduction state while the lowest-energy conduction electrons has lower probability do decay in other states. Furthermore, for the state at bottom in the LBC the states closest in energy are those in the HVB, which are very far in energy, so that the scattering rate is negligible}. We also identified two points along the $\Lambda$ direction for which the corrections are the strongest in the monolayer coinciding with the two maxima in the LCB. In general, smaller QP correction in the trilayer can be associated with the increased number of competing scattering channels---for example, the valley along the $\Lambda$ is split into three valleys and also a larger number of phonon modes contributes (see next session).

\subsection{Phonon-modes contribution}~\label{sec:phonon}

Finally, we decompose the QP corrections onto the phonon-modes. The mono- and trilayer of \mos{} have respectively 9 and 27 phonon modes which are described in the SM. For both systems, across the BZ, we consider, $\Sigma^{\lambda}_{n\kk}(\go=\epsilon_{n\kk},T)$---the projection on each phonon mode $\lambda$ of the self-energy~\footnote{In practice, we compute $\Sigma^{\lambda}_{n\kk}(\go,T)$ in $\kk$-space from Eq.~\eqref{eq:sec_theory_9} but omitting the sum over $\lambda$}---with $n$ being the LCB. The real part of $\Sigma^{\lambda}_{n\kk}(\epsilon_{n\kk},T)$ describes the energy shift of a Bloch electron in state ($n, \kk $) and it includes the virtual scattering of the electron with a phonon of with frequency $\omega_{\lambda}$ into an intermediate state ($n',\kk - \qq$) (Fan term) and the change in average potential experienced by the electrons, due to the smearing of the crystal potential induced by thermal vibration (Debye-Waller term). The imaginary part (results shown in the SM) carries the physical meaning of a finite electron lifetime and define the decay rate to its original state by re-emitting phonon.

\begin{figure}
\includegraphics[width=1.05\linewidth]{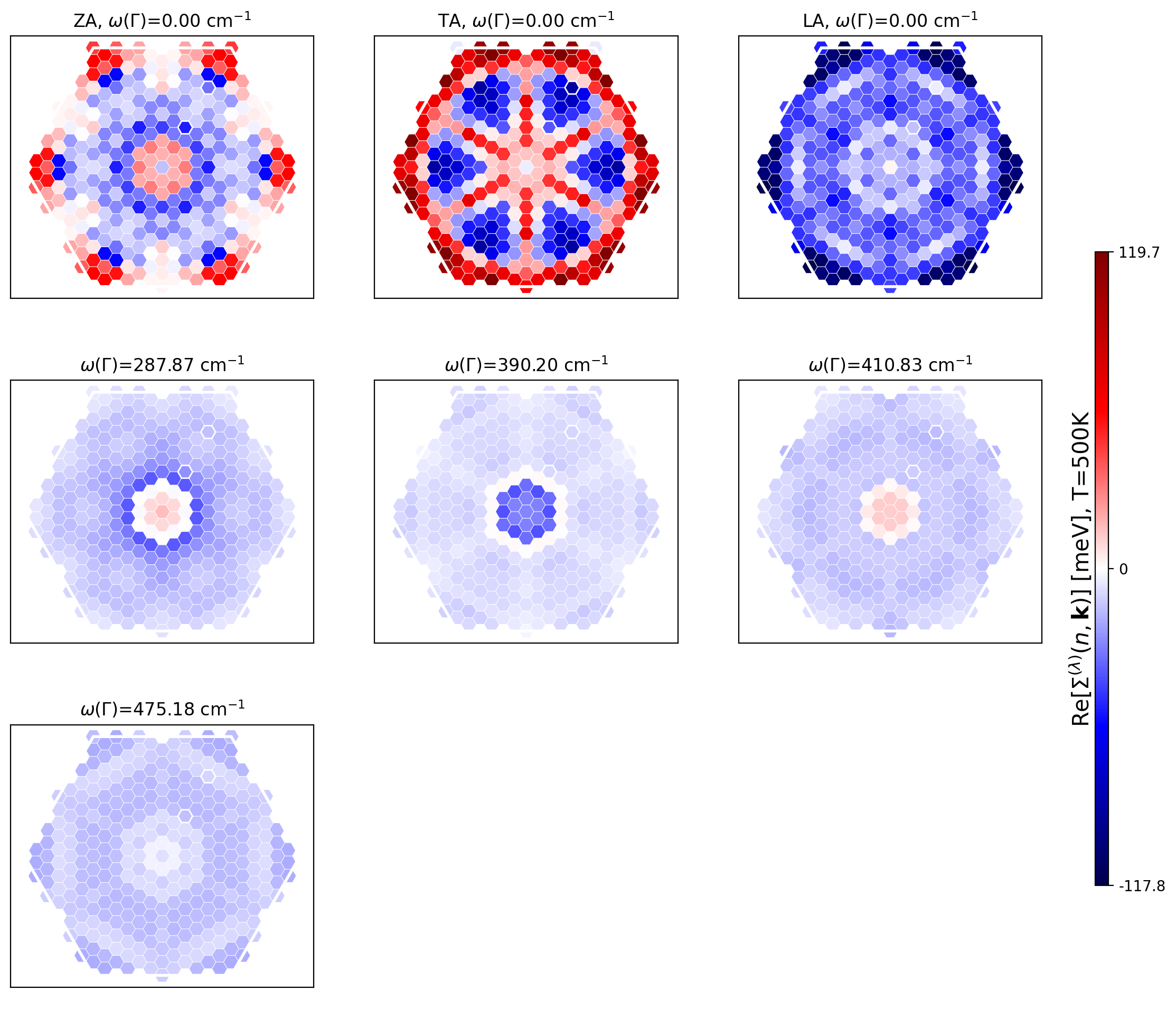}
    \caption{Real part of $\Sigma^{(\lambda)}_{n\kk}(\epsilon_{n\kk},T)$ across the BZ (see text) for the phonon modes of monolayer \mos{} at 500K, with $n$ being the LCB. Degenerate optical modes are summed together.}
    \label{fig:Mono elph}
\end{figure}

For the monolayer, Fig.~\ref{fig:Mono elph} shows that the transverse and longitudinal acoustic modes (TA and LA) contribute the most to the real part of the the QP corrections of the LCB with the TA branch being primarily responsible for the $\kk$-dependence observed in Fig.~\ref{fig:QP_vs_T} (a) and it has largest contributions at the $\Lambda_{1,2}$ points for which the QP shifts are the largest in Fig.~\ref{fig:QP_vs_T} (c). The same modes contributes to the imaginary part though in that case the LA shows the strongest $\kk$-dependence (see SM). The TA and LA are also the dominant contributions at $T=0$ (see SM) though with temperature they become even stronger with respect to the optical modes.
Our results are consistent with the mode-resolved tunneling measurements in Ref.~\cite{Lee2021-zo}, which identify the TA and LA at about 1/3 along $\Gamma-K$ (consistent with $\Lambda_1$) and  LA($M/K$) among the leading acoustic phonon excitations. 

The dominant role of the acoustic modes is the phonon-branch dependence of the self-energy and specifically in the renormalization factor in front of Fan term [Eq.~\eqref{eq:Fan}]. The low-frequency typical of acoustic modes enhances the phonon displacement and the \ep coupling. The contribution of the TA to the \ep scattering was already observed in Ref.~\cite{PhysRevB.85.115317}  that shows the in-plane shear deformation of the lattice  significantly modifies the local Mo–S bond geometry and the microscopic crystal potential.
Among acoustic mode, the flexural acoustic mode (ZA)---which corresponds predominantly to out-of-plane bending motions---has a small contribution, 
consistently with the fact that the ZA mode weakly deform the electrostatic potential~\cite{Fivaz1967} and therefore provide a negligible tunneling-scattering as observed in mode-resolved tunneling~\cite{Lee2021-zo}.

\begin{figure} 
    \includegraphics[width=1.05\linewidth]{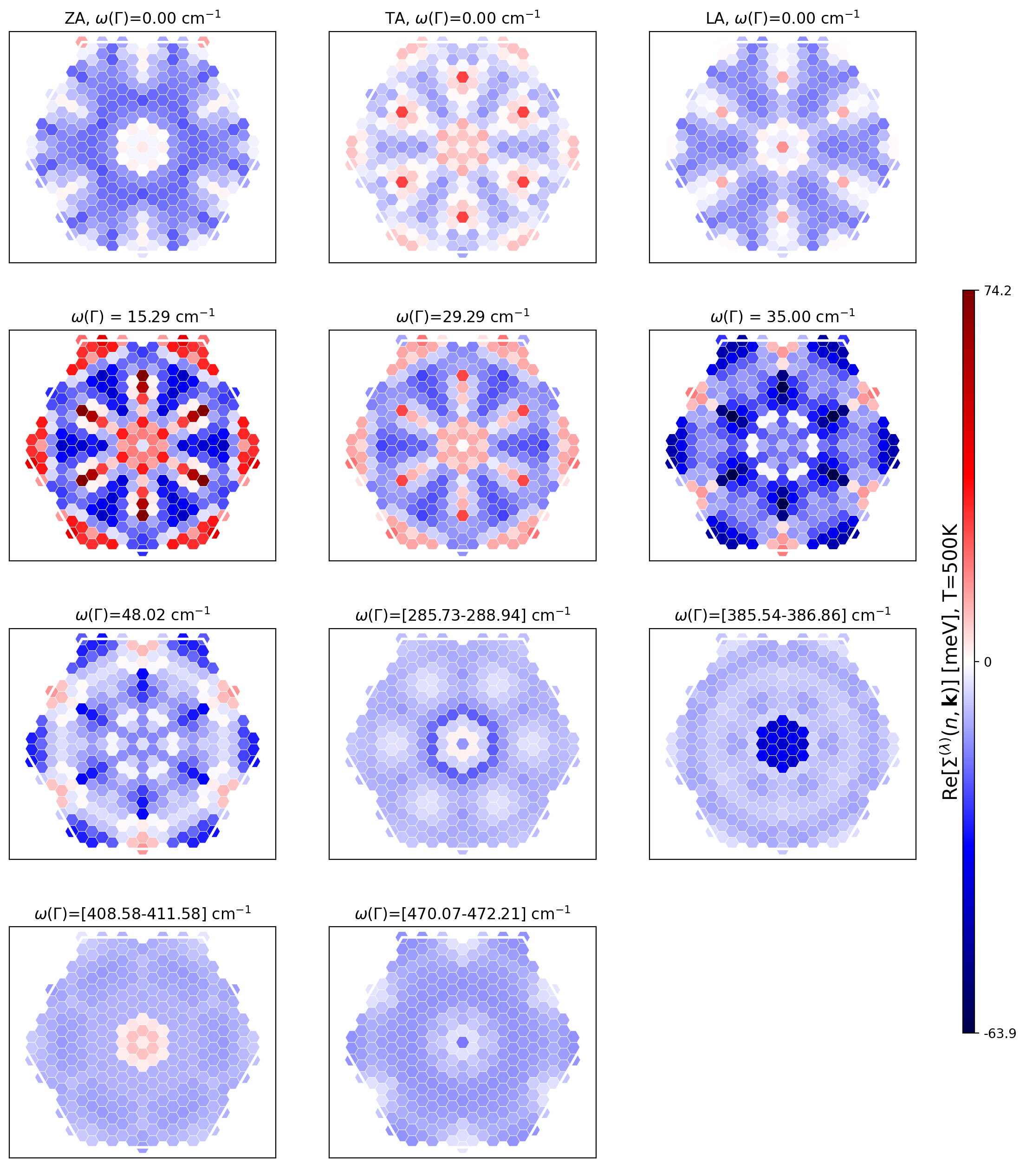}
    \caption {Real part of ${\Sigma^{(\lambda)}_{n\kk}(\epsilon_{n\kk},T)}$ across the BZ for the phonon modes of trilayer \mos{} at 500K, with $n$ being the LCB. Near-degenerate optical modes are summed together.}
    \label{fig:Tri elph}
\end{figure}

For the trilayer, Fig.~\ref{fig:Tri elph}, the strongest contributions to the QP energy shifts are associated with low-frequency optical shear modes (followed by breathing modes) originating from the splitting and hybridization of the monolayer acoustic branches in the multilayer structure~\cite{PhysRevB.84.155413}. In particular, for the lower shear mode (15.29 cm$^{-1}$), $\Re\Sigma^{\lambda}_{n\kk}$ has the most negative value (corresponding to the strongest QP correction) for the $\kk$-points along $\Lambda$ corresponding to the conduction maxima. These modes are the strongest also at $T=0$ (see SM). Consistently with the almost uniform broadening observed in Fig.~\ref{fig:QP_vs_T} across the BZ, the imaginary part of $\Sigma^{\lambda}_{n\kk}$ shows only a weak $\kk$-dependence (see SM). The shear modes contribution is large in virtue of their low-frequency and---similarly to the TA for the monolayer---for the shear deformation of the lattice which modifies the local Mo–S bond geometry (and in turn affect strongly the electronic structure along the $\Lambda$ direction). At difference with the acoustic modes, shear modes give an interlayer, rather than in-plane deformation and therefore couples strongly with the interlayer states along the $\Lambda$ direction. Breathing modes, at difference with the ZA in the monolayer, also couples relatively strongly. This can be understood~\cite{Fivaz1967} from the large anisotropy in the out-of-plane direction of layered structures which lead to large deformation potential.

\subsection{Extraordinary T-dependence of the SHG}
\label{sec:discussion}
We discuss here the uncharacteristic increase of SHG with temperature in monolayer \mos{} observed in Ref.~\cite{khan2020extraordinary} at 900 nm. First, we review the hypothesis, put forward by the authors, that the increase is due to the T-induced increase of the effective layer thickness (chalcogen planes distance). The thickness increase is assumed to be as large as 4\% since the planar expansion is limited by the small expansion coefficient of the substrate and a mismatch strain builds up and the excess energy is transmitted to the out-of-plane direction. 

\begin{figure}
    \centering
    \includegraphics[width=0.9\linewidth]{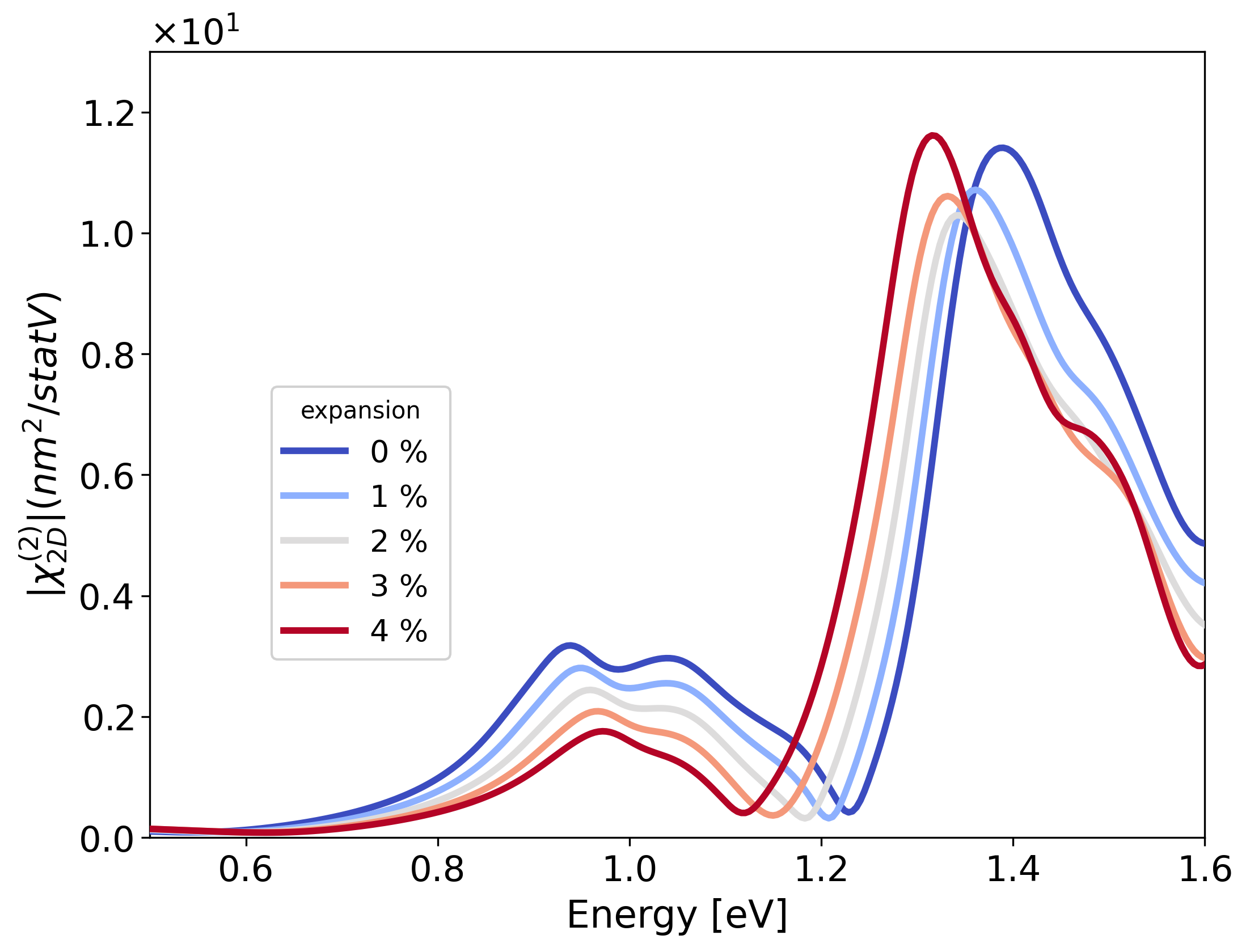}
    \caption{The surface SHG of monolayer \mos{} at various stretching of the distance between the S and Mo planes, $z_S$. Calculations at no-T level with a phenomenological broadening of 0.1~eV.}
    \label{fig:chi_2 a d exp}
\end{figure}

In Fig.~\ref{fig:chi_2 a d exp}, we consider 1--4\% increase of the out-of-plane position of the chalcogen, $z_S$ [panel (b)]. The increase of $z_S$, beside reducing the SHG in the A region, slightly shifts the C peak. We observe an increase in the SHG at 4\% for the C peak though much more modest than that predicted in Ref.~\cite{khan2020extraordinary}. We also observe that for smaller values of the stretching the SHG is reduced. The intensity of the C peak is mostly due to van-Hove peak singularities along $\Lambda$. These singularities are particularly strong due to the alignment of the HVB and LCB (nesting). Changes in the Mo-S bonding change the nesting and thus the intensity of van-Hove peak singularities in a way that is not necessarily monotonic with the stretching (temperature)---that is what we observe in Fig.~\ref{fig:chi_2 a d exp} (b). Such non-monotonic behavior of the SHG with stretching does not agree with what observed experimentally for increasing temperature. Further, in Appendix~\ref{sec:estimate}, from the anharmonic part of phonon-shift of Raman modes~\cite{Singh2023}, we roughly estimated a 0.4\% increase of the effective layer thickness at 500K---consistent with the Mo-S bond length stretching with temperature measured in bulk~\cite{Caramazza2016}. An accurate and consistent evaluation of the effective thickness expansion would require the simultaneous treatment of thermal expansion and anharmonic lattice fluctuations from first-principles while accounting for the effect of the substrate on the planar lattice expansion, which is outside the scope of this work. On the other hand, according to the results and analysis above we can offer {\em an alternative explanation based on the \ep coupling}.

We have found that electron--phonon coupling produces a general redshift and broadening of the optical structures
which reduces the amplitude of the principal spectral features. Nevertheless, the spectral weight is also redistributed by the quasiparticle-energy renormalization, and local increases of the SHG response occur in restricted energy intervals. Therefore, an increase of the signal at a fixed photon energy does not imply an enhancement of the SHG across the whole spectrum. Then, a direct quantitative comparison with Ref.~\cite{khan2020extraordinary} is particularly sensitive to the spectral position of the probe. The excitation wavelength of $900$~nm corresponds to an incoming photon energy of about $1.38$ eV that coincides with the C peak in Fig.~\ref{fig:chi_2} (a). The experimental samples were supported on SiO$_2$/Si, whereas the calculations presented here refer to free-standing \mos{} which makes a direct comparison difficult. In fact, it has been shown experimentally that the SHG on supported samples changes depending on the substrate and the sample processing~\cite{10.1021/acs.jpcc.6c01491}. In particular, it has been shown that for  SiO$_2$/Si the C peak is shifted at around 800 nm (1.55 eV). Then, measurements at 900 nm in Ref.~\cite{khan2020extraordinary} could be at ~0.2 eV below the C peak, which would coincide with the region at which we observe an increase of the SHG in Fig.~\ref{fig:chi_2} (a). If confirmed, this would explain the observed increase of SHG in monolayer \mos{} as a consequence of the strong \ep coupling for states along the $\Lambda$ valleys with TA phonons. 

In the case of the trilayer, according to our results, spectral features are red-shifted by about 0.1~eV with respect to the monolayer, so that the same excitation wavelength does not necessarily probe equivalent spectral features in the two systems. This could explain the opposite temperature dependence of the SHG observed experimentally for the monolayer and few-layer systems.

Ultimately, a definitive explanation would require on the experimental side, wavelength-resolved temperature-dependent measurements so to observe the evolution of the spectral features; on the theoretical and computational side the inclusion of missing effects that may be relevant: excitonic effects, the temperature-dependent geometry (so phonon anharmonic effects) and substrate-induced strain\footnote{Temperature-dependent Raman studies show that supported and suspended MoS$_2$ exhibit substantially different thermal evolution because the mismatch between the thermal-expansion coefficients of MoS$_2$ and the substrate generates temperature-dependent strain. The substrate can also modify the electronic structure through dielectric screening, interfacial charge transfer and adsorbates. These effects may shift the optical resonances relative to the free-standing calculation and thereby change the sign of the SHG variation measured at a fixed photon energy.}.

\section{Conclusions}
We put forward a first-principles real-time approach which includes the effect of \ep interactions on quasiparticle energies and dephasing which allows one to calculate nonlinear optical properties at finite temperature. The \ep renormalization of the quasiparticle energies and lifetimes are evaluated according to the approximated expressions for the electronic self-energy routinely used in first-principles \ep calculations~\cite{giustino2017electron}. Within the linear-response limit, the approach is shown to be formally equivalent to the finite-temperature linear-response formulation at the independent (quasi)particle level. When excitonic effects are included into the effective Hamiltonian, our treatment would introduce an artificial dephasing channel for the lowest exciton, so that our approach is (formally) limited to independent (quasi)particles. 

Within this approach, we have calculated the SHG in monolayer and trilayer \mos{}. We observe that, because of resonances at both one- and two-photons, the interpretation of the SHG is more complex of the absorption spectrum, especially close to the absorption onset. The spectral weight is overall reduced with increasing temperature but it is also redistributed, so that, in some frequency region we do observe an increase of the SHG with increasing temperature as reported in Ref.~\cite{khan2020extraordinary}, though missing effects (substrate, excitons, thermal expansion and anharmonicity) do not allow for a quantitative comparison. 

The nontrivial behavior of the SHG spectra with temperature is due to the strong crystal-momentum dependence of the quasiparticle corrections, which in turn is due to the strong crystal-momentum dependence of the coupling to the dominant phonon modes---acoustic in the monolayer and shear modes in the trilayer. The reason of such strong crystal-momentum dependence lies ultimately in the valley conformation of the BZ, which favors scattering processes at saddle points along the $\Gamma-K$ high-symmetry direction. 

Though the example in this work focuses on the SHG, the proposed real-time approach can be used to obtain any nonlinear optical property (see e.g. Refs.~\cite{mao2025shift, 10.21468/SciPostPhys.19.5.129, attaccalite2018two}) and its temperature behavior.

Coming back to the limitations of the approach, an extension to correctly include excitonic effects, would require to relax the static approximation of the Fan self-energy---which nevertheless would make the approach computationally more expensive. Regarding the underlying harmonic approximation, one could include phonon anharmonic effects in the electron-phonon lifetime---which has been shown important in anharmonic crystals~\cite{Kong2026-no}---by following the workflow recently proposed in Ref.\cite{Kong2026-ad}.  

\begin{acknowledgments}
                A.R. and M.G. acknowledge funding from the UKRI Horizon Europe Guarantee funding scheme (EP/Y032659/1). C.A. acknowledges funding from European Research Council MSCA-ITN TIMES under grant agreement 101118915. The authors are grateful for the use of the computing resources from the Northern Ireland High Performance Computing (NI-HPC) service funded by EPSRC (EP/T022175); and the ARCHER2 UK National Supercomputing Service (https://www.archer2.ac.uk)~\cite{beckett2024archer2} through the UKCP consortium funded by EPSRC (EP/X035891/1). 
                C.A. acknowledges B. Demoulin and  A. Saul for the management of the computer cluster \emph{Claudia}.  The authors thank Lorenzo Stella and Tchavdar Todorov for useful discussion on the method and results analysis. 
\end{acknowledgments}

\appendix
\section{Linear-response limit of the Berry's polarization}
\label{linearP}
The Berry polarization along the cartesian direction $\alpha$ is given by,
\begin{equation}\label{eq:pberry}
P_\alpha(t)=\frac{ef}{(2\pi)^3}\sum_{m}\int_{BZ}d\kk\, \Im\langle v_{m\kk}(t)|\partial_{{\kk}_\alpha} v_{m\kk}(t)\rangle,
\end{equation}
where $f$ is the spin degeneracy factor and $e$ the electron charge.
Within linear response, the constitutive relation between the polarization and the electric field reads,
\begin{equation}\label{eq:chiE}
    P_\alpha^{(1)}(t) = \int dt' \chi^{(1)}_\alpha(t-t') \efield_{\alpha}(t').
\end{equation}
In what follows, starting from a perturbative expansion of Eq.~\eqref{eq:pberry} in the power of the electric field ${\bf \efield}$ we find the usual expression for ${\bf P}^{(1)}$ in terms of the density matrix $\rho_{cv,\kk}$ and the dipole matrix elements $d_{cv\kk}^\alpha$,
\begin{equation} \label{eq:P1_fin}
    P_\alpha^{(1)}(t) = -\frac{1}{(2\pi)^3} \int d\kk\,\sum_{cv} 2\Re\left(\rho^{}_{cv,\kk} d_{cv{\kk}}^{\alpha}\right),  
\end{equation}
where the $v,c$ are the index for the valence and conduction bands respectively. 

To the first order in the electric field $E_\alpha$, the time-dependent valence states,
\begin{equation}
|v_{m\kk}(t)\rangle \approx |v_{m\kk}^{(0)}\rangle + |v_{m\kk}^{(1)}(t)\rangle
\end{equation}
where $|v_{m\kk}^{(0)}\rangle$ is the unperturbed ground-state orbital and $|v_{m\kk}^{(1)}(t)\rangle$ is the first-order time-dependent variation. Substituting this into the Eq.~\eqref{eq:pberry} and keeping only terms linear in the perturbation, we obtain,
\begin{align}\label{eq:P1_sub1}
&P_\alpha^{(1)}(t)=\\&\frac{ef}{(2\pi)^3}\sum_{m}\int d\kk~\Im[\langle v_{m\kk}^{(0)}|\partial_{\kk_\alpha}v_{m\kk}^{(1)}(t)\rangle+\langle v_{m\kk}^{(1)}(t)|\partial_{\kk_\alpha}v_{m\kk}^{(0)}\rangle] \notag
\end{align}
When we perturb a system from its fully occupied valence manifold, the physical variation exists purely in the empty conduction bands. This imposes the gauge condition $\langle v_{m\kk}^{(0)}|v_{m\kk}^{(1)}(t)\rangle=0$, when $m$ is a valence state. 
By taking the $\kk$-derivative of this orthogonality condition, we find:
\begin{align}\label{eq:devkGC}
&\partial_{\kk_\alpha}\langle v_{m\kk}^{(0)}|v_{m\kk}^{(1)}(t)\rangle=0 \Rightarrow \\&\langle v_{m\kk}^{(0)}|\partial_{\kk_\alpha}v_{m\kk}^{(1)}(t)\rangle =-\left(\langle v_{m\kk}^{(1)}(t)|\partial_{k}v_{m\kk}^{(0)}\rangle\right)^{*}\notag
\end{align}
Using this condition into Eq.\eqref{eq:P1_sub1} and noticing that the integral takes the form $Z-Z^{*}$, which simplifies to $2i \, Im(Z)$, we obtain,
\begin{equation}\label{eq:P1_sub2}
P^{(1)}_\alpha(t)=\frac{2ief}{(2\pi)^3}\sum_{m}\int dk\,\Im\langle v_{m\kk}^{(1)}(t)|\partial_{k} v_{m\kk}^{(0)}\rangle.
\end{equation}
We thus project the first-order variation onto the unperturbed conduction bands manifold $|v_{c\kk}^{(0)}\rangle$,
\begin{equation}
	|v_{m\kk}^{(1)}(t)\rangle = \sum_{c}a_{cm\kk}^{(1)}(t)|v_{c\kk}^{(0)}\rangle,
\end{equation}
where $m$ is a valence state.
Substituting this expansion into Eq.~\eqref{eq:P1_sub2}, the linear response expression for the polarization becomes:
\begin{equation}\label{eq:P1_sub3}
P^{(1)}_\alpha(t)=-\frac{ef}{\pi}\sum_{v,c}\int_\text{BZ}d\kk~2Re\left(a_{cv,\kk}^{(1)}(t)\langle v_{v\kk}^{(0)}|i\partial_{\kk_\alpha}|v_{c\kk}^{(0)}\rangle\right).
\end{equation}
The term $e\cdot\langle v_{v\kk}^{(0)}|i\partial_{\kk_\alpha}|v_{c\kk}^{(0)}\rangle$ is the interband dipole matrix element $d^\alpha_{cv,\kk}$. 
The off-diagonal term of the density matrix in the linear response limit writes,
\begin{equation}
\rho_{cv\kk}^{(1)} = 
f\sum_{m}^\text{occ} (a_{mc,\kk}^{(0)} + a_{mc,\kk}^{(1)}) (a_{mv,\kk}^{(0)} + a_{mv,\kk}^{(1)})^* = f\, a_{cv,\kk}^{(1)},
\end{equation}
since in the unperturbed system $a_{mi,\kk}^{(0)} = \delta_{mi}$ (where $m$ is a valence state) and the above orthogonality requirement for the first-order variation $a_{mc,\kk}^{(1)}$ are the only non-zero terms. Then, substituting into Eq.~\eqref{eq:P1_sub3}, we arrive at Eq.~\eqref{eq:P1_fin} which is used in Sec.~\ref{sec:linlim}.
Finally, the linear-response limit of the Berry-phase polarization, is reflected also in the expression of the field-system coupling operator $U_\kk$ that takes the usual form (thus valid within linear response) $\efield\cdot\hat {\bf d}_\kk$:
\begin{equation}
    \hat U^{(1)}_{\kk,\alpha} \ket{v^{(1)}_{v\kk}} \propto \efield_\alpha \frac{\delta P^{(1)}_\alpha}{\bra{\delta v^{(0)}_{v\kk} }} = \efield_\alpha \sum_c d^\alpha_{cv,\kk} \ket{v^{(1)}_{v\kk}}.
\end{equation}
Then, in the Kohn-Sham basis representation,
\begin{equation}
    \hat U^{(1)}_{vc \kk,\alpha} = \efield_\alpha d^\alpha_{vc,\kk}.
\end{equation}
\section{Double-grid integration method for the \ep self-energy} \label{sec:double grid}
The double-grid integration method is an approximate approach to evaluate the integral on Brillouin space of the self-energy. This approach has been applied in the solution of the Bethe-Salpeter equation\cite{kammerlander2012speeding,10.3389/fchem.2021.763946} and in the calculation of the \ep self-energies.\cite{brunin2020phonon,chaves2020boosting}

In the quasi-particle approximation, the \ep contribution to the self-energy has the general structure,
\be \label{eq:genEPH}
\Sigma_{\nk}\(\go,T\) = \frac{1}{N_q}\sum_{n'\gql}  \frac{ D_{\qq\gl n'}\(T\)}{\gee_{n \kk}-\gee_{n' \kk-\qq} \pm \goql -i0^{+}},
\ee
where $ D_{\qq\gl n'}\(T\)$ is the numerator as the one in Eq.~\eqref{eq:Fan}. The key aspect is that the denominator varies much faster in the Brillouin space than the numerator. 
As well, the terms in the denominators are electronic and the phonon energies---which we show below can be obtained inexpensively from interpolation---whereas the numerator requires the calculation of the \ep matrix elements and it is therefore computationally more expensive. From here the idea to calculate the numerator and denominator on two distinct grids in the Brillouin zone: a finer grid for the fast-varying, relatively computationally inexpensive denominator and a coarser grid for the slow-varying computationally intensive numerator. Using this idea, we rewrite the self-energy in Eq.~\ref{eq:genEPH} as,
\begin{align}
&\Sigma_{\nk}\(\go,T\) = \frac{1}{N_q}\sum_{n'\gql}  D_{\qq\gl n'}\(T\) \times \\ &\frac{1}{N_{\tilde \qq}}\sum_{\tilde \qq \in \qq} \frac{1}{\gee_{n \kk}-\gee_{n' \kk-\tilde \qq} \pm \goqlt -i0^{+}}, \notag
\end{align}
where $\tilde \qq$ are the $\qq$-points of the finer grid in the vicinity of a $\qq$ point in the coarse grid. As anticipated, the electronic and the phonon energies on the finer grid are obtained by interpolation.
Given a regular $\qq$-grid, we calculate the real-space interatomic force constants (IFCs) via discrete (fast) Fourier transforms. Then the IFCs thus obtained can be used to calculate inexpensively via (inverse) Fourier transform dynamical matrices at any $\tilde \qq$ vector not included in the coarse reciprocal-space mesh.\cite{baroni2001phonons} The electronic energies $\gee_{n' \kk-\tilde \qq}$ are calculated using a smooth Fourier interpolation.~\cite{pickett1988smooth}

\section{Estimate of $z_S$ dependence with T} 
\label{sec:estimate}
As discussed in Sec.~\ref{sec:discussion}, in a supported monolayer---where thermal expansion is constrained by the substrate---temperature mostly affects the distance $z_S$ of the chalcogens from the Mo plane. According to various studies of the temperature-induced phonon shift, this is mostly driven by anharmonic effects and specifically by three- and four-photon scattering~\cite{Najmaei2013,Lanzillo2013,Taube2014,Singh2023}. In Ref.~\cite{Singh2023}, by analyzing the phonon shift of the Raman active $A_{1g}$ and $E^1_{2g}$, they extracted the anharmonic part $\Delta \omega_\text{an}$ and decomposed it into three- and four-photon contribution. Here, we extract an estimate of the $z_S$ stretching from a simple model of a one-dimensional quantum oscillator including---beside the harmonic contribution---a cubic contribution 
\be
V(q) =  \frac12 m\omega_0^2 q^2 + \frac{\lambda \sqrt{2}  \hbar \omega_0}{d^3} q^3,
\ee
 where $d = \sqrt{\hbar/(m\omega_0)}$, with $\omega_0$ is the phonon frequency and $m$ the reduced mass corresponding of the phonon mode.
The mean displacement from equilibrium can be found by setting the forces to zero, $\bigl \langle dV/dq \bigr \rangle$,
\be
\bigl \langle q \bigr \rangle = -\frac{3\sqrt{2}\lambda}{d}  \bigl \langle q^2 \bigr \rangle,
\ee
where the mean square displacement (unperturbed) is 
\be
 \bigl \langle q^2 \bigr \rangle = \frac{1}{2} d^2 (2n + 1),
\ee
with $n$ the phonon population (depending on temperature though the Bose-Einstein distribution).
From the change in the curvature, $\bigl \langle d^2V/dq^2 \bigr \rangle$,
\be
 \bigl \langle d^2V/dq^2 \bigr \rangle = m\omega_0^2 + 6 \frac{\lambda \sqrt{2}  \hbar \omega_0}{d^3}\bigl \langle q \bigr \rangle, 
\ee
one can then extract the energy shift with respect the unperturbed system,
\be
\hbar \Delta \omega = -9 \lambda^2 \hbar\omega_0 (2n + 1).
\ee
Then, $\lambda = \frac13 \sqrt{\frac{-\Delta\omega }{\omega_0(2n+1)}} $ where $\Delta \omega$ and $\omega_0$ were extracted from Raman measurements in Ref.~\cite{Singh2023}. We can thus estimate $\lambda$ and $\bigl \langle q \bigr \rangle$ for the $A_{1g}$ and $E^1_{2g}$ modes. Combining those and considering the geometry of the system, we estimate that at 500K $z_S$ is stretched by about 0.4\%.

\end{document}